\documentclass[reprint,aps,prl,twocolumn,superscriptaddress,showpacs,preprintnumbers,amsmath,amssymb,floatfix,merge]{revtex4-1}
\usepackage[]{hyperref} 	
\usepackage[]{epsfig}
\usepackage{graphicx}
\usepackage{dcolumn}
\usepackage{epstopdf}
\usepackage{multirow}

\usepackage{color}

\newcommand{\queens}{Department of Physics, Queen's University, Kingston, ON, Canada, K7L 3N6}

\newcommand{\ensl}{Laboratoire de Physique de l'Ecole Normale Sup\'{e}rieure de Lyon, CNRS and Universit\'{e} de Lyon, France}

\newcommand{\ilm}{Institut Lumi\`ere Mati\`ere, UMR5306 Universit\'e Lyon 1-CNRS, Universit\'e de Lyon, France}

\newcommand{\insa}{Universit\'e de Lyon, INSA-Lyon, MATEIS, CNRS UMR 5510, F-69621 Villeurbanne, France}

\newcommand{\mins}{\mbox{$\mbox{min}$}}
\newcommand{\hours}{\mbox{$\mbox{h}$}}
\newcommand{\secs}{\mbox{$\mbox{s}$}}

\newcommand{\eV}{\mbox{$\mbox{eV}$}}

\newcommand{\keV}{\mbox{$\mbox{keV}$}}

\newcommand{\GeV}{\mbox{$\mbox{GeV}$}}

\newcommand{\MeV}{\mbox{$\mbox{MeV}$}}

\newcommand{\nm}{\mbox{$\mbox{nm}$}}
\newcommand{\microm}{\mbox{$\mu\mbox{m}$}}
\newcommand{\micros}{\mbox{$\mu\mbox{s}$}}

\newcommand{\ms}{\mbox{$\mbox{ms}$}}
\newcommand{\Newt}{\mbox{$\mbox{N}$}}
\newcommand{\mm}{\mbox{$\mbox{mm}$}}
\newcommand{\mK}{\mbox{$\mbox{mK}$}}

\newcommand{\msq}{\mbox{$\mbox{m}^2$}}
\newcommand{\mmsq}{\mbox{$\mbox{mm}^2$}}
\newcommand{\mmcu}{\mbox{$\mbox{mm}^3$}}

\newcommand{\kbr}{\mbox{$\mbox{KBr}$}}
\newcommand{\microJoules}{\mbox{$\mu \mbox{J}$}}

\newcommand{\Joules}{\mbox{$\mbox{J}$}}
\newcommand{\GPa}{\mbox{$\mbox{GPa}$}}
\newcommand{\bit}{\mbox{$\mbox{Bi}^{3+}$}}

\newcommand{\bgo}{\mbox{$\mbox{Bi}_4\mbox{Ge}_{3}\mbox{O}_{12}$}}
\newcommand{\cdwo}{\mbox{$\mbox{CdWO}_{4}$}}
\newcommand{\cawo}{\mbox{$\mbox{CaWO}_{4}$}}
\newcommand{\zwo}{\mbox{$\mbox{ZnWO}_{4}$}}

\newcommand{\baft}{\mbox{$\mbox{BaF}_2$}}
\newcommand{\mgo}{\mbox{$\mbox{MgO}$}}

\newcommand{\integratorSatLevel}{85}

\begin{document}

\title{Sound and light from fractures in scintillators
}

\affiliation{\queens}
\affiliation{\ilm}
\affiliation{\ensl}
\affiliation{\insa}
\author{A.~Tantot} \affiliation{\queens} \affiliation{\ilm}
\author{S.~Santucci} \affiliation{\ensl}
\author{O.~Ramos} \affiliation{\ilm}
\author{S.~Deschanel}   \affiliation{\insa} \affiliation{\queens}
\author{M.-A.~Verdier}   \affiliation{\queens}
\author{E.~Mony} \affiliation{\queens}
\author{Y.~Wei}   \affiliation{\queens}
\author{S.~Ciliberto} \affiliation{\ensl}
\author{L.~Vanel} \affiliation{\ilm}
\author{P.~C.~F.~Di~Stefano}  \email{Corresponding author: distefan@queensu.ca}  \affiliation{\queens}

\date{\today}

\begin{abstract}
Prompted  by intriguing events observed in  certain particle-physics searches for rare events, we  study light  and acoustic emission simultaneously in some inorganic scintillators subject to mechanical stress.  
We observe mechanoluminescence in  \bgo, \cdwo\ and \zwo, in various mechanical configurations at room temperature and ambient pressure. 
We analyze the temporal and amplitude correlations between the light emission and the acoustic emission during  fracture. 
A novel application of the precise energy calibration of \bgo\ provided by radioactive sources allows us to deduce that the fraction of elastic energy converted to light is at least $3 \times 10^{-5}$.
\end{abstract}


\maketitle

Rare-event searches in particle physics, like those looking for  particle dark matter~\cite{schnee_introduction_2011}, 
employ draconian measures to reduce all forms of background that could hide their signal.  The main background is usually of radioactive nature, however other forms  are possible.  
For instance, the  first, calorimetric phase of the CRESST experiment  measured phonons in sapphire crystals cooled to around $10~\mK$ to detect particle interactions~\cite{angloher_limits_2002}.  
Initially, thermal contraction of the holders caused fractures in the crystals, limiting sensitivity of the experiment by three orders of magnitude in rate above the radioactive background until the problem was identified~\cite{astrom_fracture_2006,stefano_cresst_2001}.
Certain more recent experiments use a coupled measurement of phonons and scintillation  to determine the nature of the interacting particles and reject the radioactive background~\cite{angloher_results_2011}.
However, mechanoluminescence, i.e. the emission of light by a solid subject to mechanical stress, 
occurs in many materials~\cite{zink_squeezing_1981}, including plastic scintillators~\cite{reynolds_mechanoluminescence_2000}.
In addition, correlated acoustic emission (AE) and visible-photon emission have been observed during mechanical deformation of colored alkali halides~\cite{chandra_acoustic_1984}.  
Other types of emission, including electrons~\cite{langford_simultaneous_1987}, positive ions~\cite{dickinson_electron_1985},  and X-rays~\cite{Camara2008},
have also been studied in various materials under mechanical stress.
Nevertheless, there is an opinion that fracture events in scintillation-phonon detectors would produce little or no light (e.g.~\cite{gaitskell_towards_2001}).

Prompted by  recent intriguing events observed by the CRESST~II dark matter search using cryogenic scintillation-phonon \cawo\ detectors~\cite{angloher_results_2011}, 
we study mechanoluminescence as a form of photon-producing background in such devices.
We also argue that photon emission can help characterize the rupture dynamics in a manner 
analogous
to AE~\cite{mogi_study_1962,Garcimartin1997a,Deschanel2006a}, 
and can provide complementary information about  the fracture energy.  
To investigate these two points, we have carried out what we believe are the first experiments to correlate the acoustic and light emission from common scintillators under mechanical stress, and to quantify the conversion of elastic energy into light.

\begin{figure}[h]
	\centering
	\epsfig{file=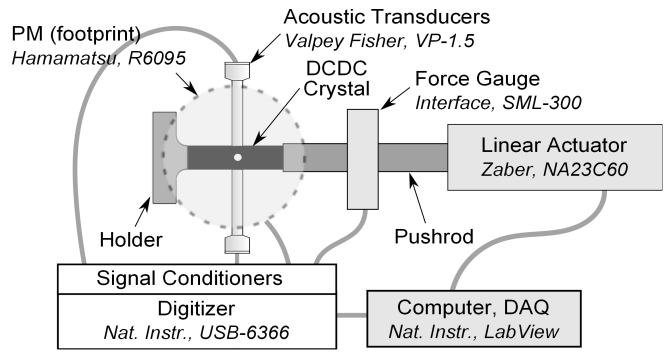,width=\linewidth}
	\caption[Fig_1]{Actuator-driven experimental setup with DCDC sample geometry.}
	\label{fig_Fig_1}
\end{figure}

Our studies involve the measurement of acoustic phonons and luminescence as a piece of scintillating crystal is stressed to the point of rupture.  
The inorganic scintillators studied here are widely used in particle detection~\cite{knoll_radiation_2000,rodnyi_physical_1997}: bismuth germanate (\bgo, a.k.a. BGO), zinc tungstate (\zwo)  
and cadmium tungstate (\cdwo).  Except where specified, all samples were kindly supplied by Crystal Manufacturing~Lab~Ltd., Novosibirsk.
We have tested two loading geometries, at room temperature and ambient pressure.
Initially, $20 \times 10 \times 5~\mmcu$ rectangular prisms were indented by a steel bead driven by a manual screw, causing the material to break in multiple fragments.
In order to better control the rupture process, we subsequently adopted the double cleavage drilled compression (DCDC) geometry~\cite{janssen_specimen_1974} (Fig.~\ref{fig_Fig_1}).  
These samples were $20 \times 5 \times 3~\mmcu$ rectangular prisms, polished to optical quality, with a $1~\mm$ diameter circular hole drilled perpendicularly in the middle of the $20 \times 5~\mmsq$ face. 
Under compression along the long axis, cracks formed reproducibly on either side of the hole in the plane parallel to the $20 \times 3~\mmsq$ face.
Unlike BGO, tungstates have a cleavage plane~\cite{danevich_crystals_2005}; orientation of the crystals was chosen so that this plane was parallel to the DCDC fracture direction. 
The DCDC samples were pressed against a backstop by a pushrod that was  driven by an actuator (controlled manually in early experiments, and by computer in later ones).
This geometry provides some control of crack velocity via the applied compressive stress, up to a critical length at which point the sample cleaves abruptly~\cite{janssen_specimen_1974}.
A force gauge in the pushrod measured the load imposed by displacements of the 
actuator. 
The acoustic activity was recorded by two piezo-electric pinducers in contact with opposite sides of the crystal via ultrasound gel. 
Light emission was recorded by a standard bialkali photomultiplier (PM) covering the crystal.  
The output of the PM was integrated 
with a time constant ranging from $1.8~\micros$ to $10~\micros$ depending on the sample
to produce a signal that could be digitized more slowly than the raw PM output. 
All four channels (force, 2 AE, PM) were digitized at a rate of $2 \times 10^6$ samples per second in a continuous stream.
The setpoint of the computer-controlled actuator was also recorded.

\begin{figure}[h!]
	\centering
	\epsfig{file=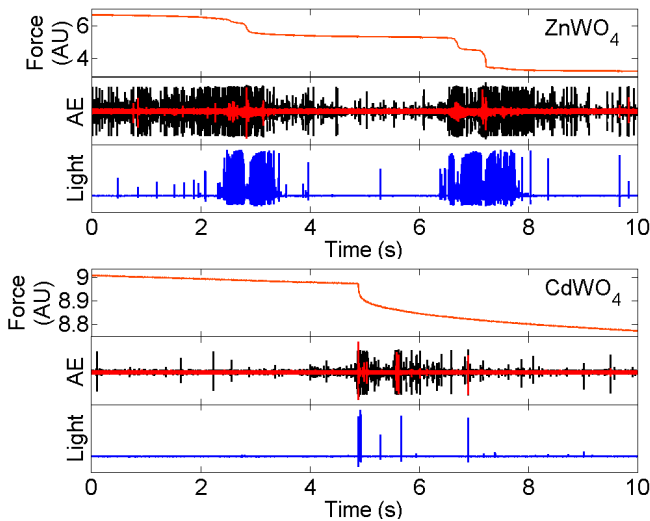,width=\linewidth}
	\caption[Fig_1]{Ten-second samples from observations of manually 	compressed DCDC specimens of \cdwo\ and \zwo. In each figure, top to bottom: applied force, both AE signals, and light signal. Drops in the force indicating fracture are correlated to increased acoustic and photon activity. All signals are uncalibrated.}
	\label{fig_MainResult}
\end{figure}

The fracture procedure involves advancing the actuator by small steps (down to $0.1~\microm$). Each step is followed by a waiting time of at least $30~\secs$.  The whole procedure lasts  $1$--$2~\hours$ during which we continuously  monitor  AE and PM activity. Figure~\ref{fig_MainResult} shows typical observations for the  tungstate crystals. %
Spurts of activity on the AE and PM channels are correlated to drops in the force.  
These abrupt force relaxations are a direct consequence of crack growth in the samples.
There is also a regime in which the force decreases slowly, and during which
other processes, such as rearrangements and dislocations, produce AE and PM signals~\cite{chandra_acoustic_1984}.
Similar correlations have been observed for other materials, such as colored \kbr~\cite{chandra_acoustic_1984}, composites~\cite{crasto_correlation_1987}, and silica glass~\cite{kawaguchi_time-resolved_1995}.

\begin{figure}[h]
	\centering
	\epsfig{file=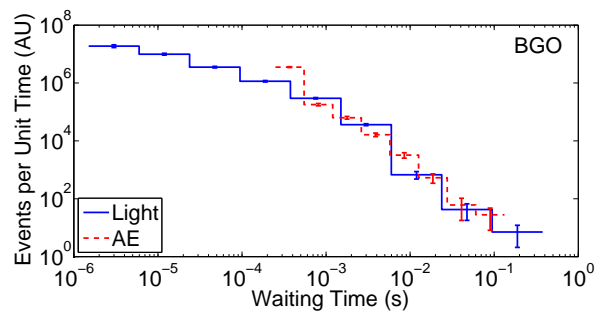,width=\linewidth}
	\caption[Waiting Times]{Waiting times between light events and between AE events during the indentation of a BGO sample.  Both channels show similar behavior over the shared part of the range, with the shorter light pulses providing better timing separation than the longer AE ones.}
	\label{fig_Wait_extra}
\end{figure}

Indentation of BGO samples (from Fibercryst, France) 
also produces correlated AE and light signals. These correlations can be analyzed more quantitatively by looking at the statistical properties of individual acoustic and light events. 
In this analysis, individual events have been determined on each channel by requiring that the signal surpass a threshold, and that it be separated from the previous event by at least a minimum time.
Figure~\ref{fig_Wait_extra} shows that the distributions of waiting times between events are  similar for AE and light over two decades. For light events, it is possible to obtain a much smaller minimum separation time ($1.5~\micros$) than for AE ($250~\micros$), since the former pulses are shorter than the latter. 

\begin{figure}[h]
	\centering
	\epsfig{file=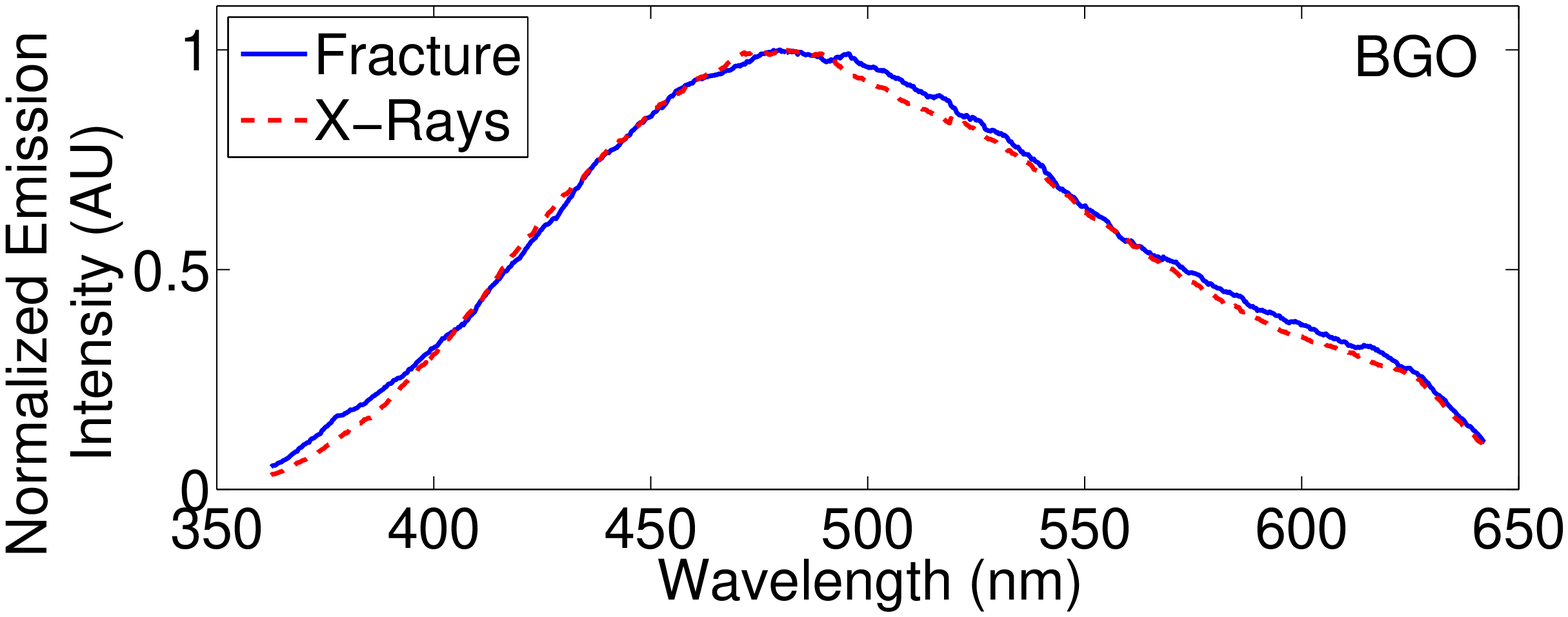,width=\linewidth}
	\caption[Fig_1]{Emission spectra from a BGO sample, first during fracture, then under X-ray excitation.  Similarity of spectra indicates a common emission mechanism.}
	\label{fig_Spectra}
\end{figure}

To  quantify the amount of light emitted and to investigate the emission mechanism, we next compared the spectra of light emitted by BGO (sample from Fibercryst) during fracture and during scintillation stimulated by X-rays.  To avoid afterglow from the high dose of X-rays, the fracture spectrum was measured first, and then one of the fragments was measured while being irradiated by X-rays.
The measurements of the light spectra were carried out with a 
monochromator and a 
CCD camera.  In the X-ray excitation configuration, 
a generator provided X-rays with a broad spectrum up to a few tens of~\keV.
Over the recorded range of wavelengths, both resulting spectra (Fig.~\ref{fig_Spectra}) show a very similar shape (as is the case for many materials~\cite{duignan_triboluminescence_2002}), indicating a common underlying luminescence mechanism, at least for the last stages of light emission.  
The likeness of our spectra to previous work on BGO indicates this last stage involves ${}^3P_1 \rightarrow {}^1S_0$  electronic transitions of the \bit\ luminescence centers~\cite{weber_luminescence_1973}. 
This measurement does not determine if the fractures excite the luminescence centers directly.
It cannot be excluded that the crystal is indirectly excited by particles produced by fracto-emission~\cite{langford_simultaneous_1987} interacting in the scintillator.  
For instance, if X-rays or electrons are produced, the BGO sample would be quite efficient at detecting them itself (attenuation length  $\approx 20~\microm$ for $10~\keV$ X-rays~\cite{henke_x-ray_1993}),
as opposed to needing an external X-ray detector~\cite{Camara2008}.
Alternatively, any UV light (potentially from arcs in the atmosphere near the fracture)
might be re-absorbed in the scintillator since BGO absorbs wavelengths below $\approx 300~\nm$~\cite{rogemond_fluorescence_1985}.

\begin{figure}[h]
	\centering
	\epsfig{file=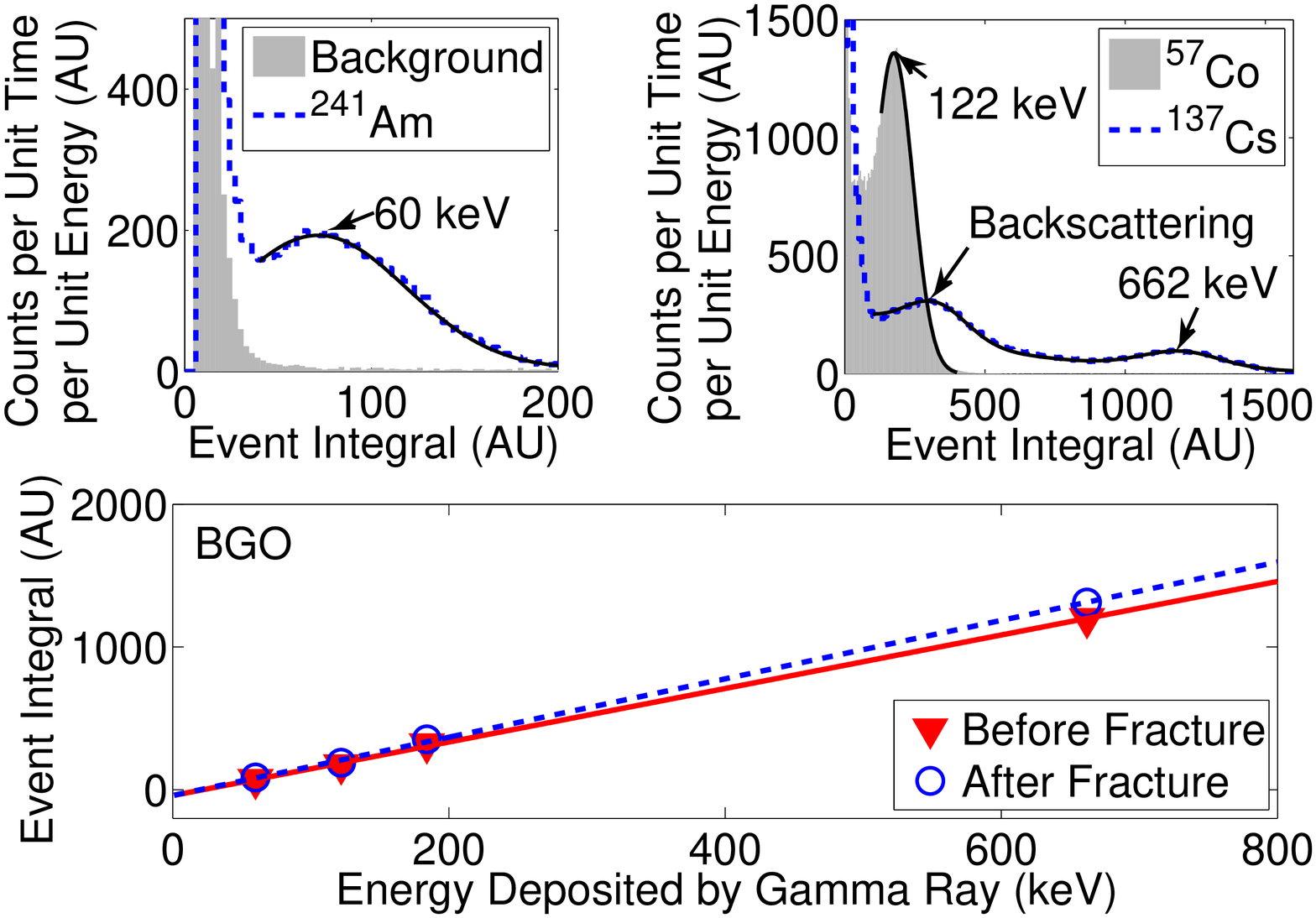,width=\linewidth}
	\caption[Calibration]{Energy calibration of a DCDC BGO scintillator using radioactive sources. Top: spectra for various sources and fits to the full-collection and backscattering peaks.  Bottom: response of system is fairly linear over this energy range, and varies by $\approx 10\%$ after the fracture.}
	\label{fig_calibr}
\end{figure}

We then calibrated the light channel for a DCDC BGO sample using radioactive $\gamma$-ray sources as we would calibrate a regular scintillator used for particle detection.  
Calibration was carried out in the main DCDC setup and with the standard DAQ (Fig.~\ref{fig_Fig_1}). 
The analysis involves extracting individual events as described earlier, and then building the distribution of the event integrals.   The integral of an event, like its amplitude, is a proxy for the energy deposited by the $\gamma$, so we can identify the peaks in the distribution corresponding to the energies of the radioactive sources and of backscattering~\cite{knoll_radiation_2000}, as   shown in Fig.~\ref{fig_calibr}.
Over the energy range available with these sources, the crystal and readout behave in a linear manner with respect to the energy deposited in the scintillator, and we extrapolate this calibration to higher energies.
Calibrations before and after the fracture are compatible to within $10\%$.
From the standpoint of the scintillation mechanism, 
$\gamma$-rays and X-rays can interact in inorganic scintillators via the photoelectric effect or through Compton scattering, creating a primary electron-hole pair which eventually transfers a portion of the deposited energy to the  luminescence centers (\bit\ for BGO).
For a given   light signal, calibrations provide the equivalent amount of energy deposited by a $\gamma$-ray.
This can be converted to the actual energy of the emitted light by knowing the light yield of the scintillator ($8~\mbox{photons}/\keV$ for $\gamma$-rays in BGO~\cite{knoll_radiation_2000}), and the energy of individual scintillation photons 
($2.6~\eV$ c.f. Fig~\ref{fig_Spectra}).

\begin{figure}[h]
	\centering
	\epsfig{file=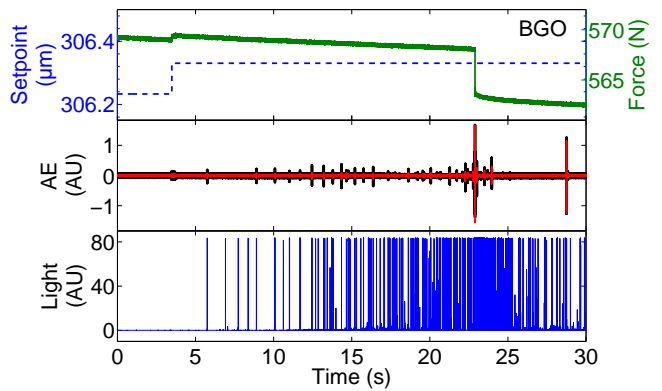,width=\linewidth}
	\caption[Calibration]{Zoom on the rupture of the DCDC BGO sample of Fig.~\ref{fig_calibr} showing correlation between the various channels when the main fracture occurs  around $23~\secs$ (corresponding to $\approx 17~\mins$ since start of procedure). 
	Photon signal shows a wide range of amplitudes (the largest of which saturate the integrator).
	}
	\label{fig_BGObreak}
\end{figure}

\begin{figure}[h]
	\centering
	\epsfig{file=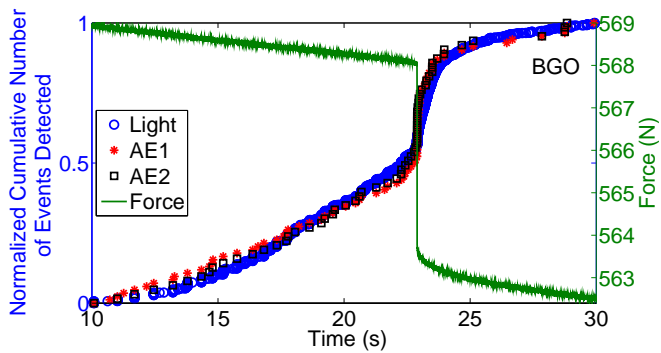,width=\linewidth}
	\caption[Normalized Cumulative Counts]{Normalized cumulative number of events on AE and photon channels around  main fracture, for  BGO sample of Fig.~\ref{fig_BGObreak}. 
	AE and photon channels show similar behavior. 
	The drop in force signalling the main fracture corresponds to a sharp increase in the event rates for photons and AE.}
	\label{fig_Corr_extra}
\end{figure}

\begin{figure}[h]
	\centering
	\epsfig{file=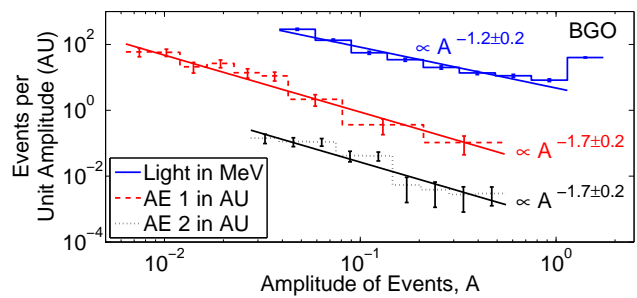,width=\linewidth}
	\caption[Amplitude spectra]{Amplitude distributions for the various channels of BGO sample from Fig.~\ref{fig_BGObreak}.  Amplitude of the light channel is proportional to energy.  All distributions are  fitted by  power laws (up to saturation, for light).}
	\label{fig_Hist_spectra}
\end{figure}

The calibrated BGO crystal was then fractured in the main DCDC setup, using the protocol described earlier.
We focus on the last drop in force, occurring when the crack splits the sample in two.
Figure~\ref{fig_BGObreak} shows that the overall behavior is similar to that observed for the tungstates.
The photon channel displays a wide distribution of event amplitudes.
They in fact reach up to  the saturation level of the integrator, equivalent to  $\gamma$-ray energy deposits of $\approx \integratorSatLevel~\MeV$ --- a testimony to the amount of light emitted. 
There is also a hint of increased photonic activity before the main fracture occurs.
The correlations are better quantified in Fig.~\ref{fig_Corr_extra} illustrating  the cumulative number of events on the acoustic channels and those on the photon channel. 
Minimum separation time required between photon events was $10~\micros$, and $1~\ms$ between  AE events.  
On all three channels, the cumulative number of events rises steadily with a slight increase in slope until the fracture, as determined by the drop  in the force.  
This drop marks a sharp increase in the slope of cumulative events that eventually peters out. 
When normalized to the total number of events on each channel, all  curves are very similar.
Lastly,  event amplitudes for all channels are power-law distributed (Fig.~\ref{fig_Hist_spectra}).  
For both  AE channels, the exponents are $-1.7 \pm 0.2$ (errors are one standard deviation).  Assuming the AE energy is proportional to the amplitude squared, the power-law distribution of the energy would have as exponent $-1.4 \pm 0.1$.  This is compatible with the exponent obtained for the light channel, $-1.2 \pm 0.2$, for which amplitude is proportional to energy.
A more precise measurement might check if the absolute value of the AE exponent is in fact larger than that of the light channel, an  indication of dissipation in the acoustic channel~\cite{olami_self-organized_1992}.

We can also use our technique to estimate certain contributions to the energy budget of the rupture process in BGO.  
Elastic energy is converted into broken bonds, phonons (that are either measured as acoustic emission or decay into heat) and light.  
For the emitted light, the calibrations with $\gamma$-ray sources  establish that the total amount of emitted light is equivalent to that caused by $\gamma$-ray energy deposits of at least $300~\GeV$, in other words a total energy converted to light greater than $6.2~\GeV$.  
Only a lower bound is set because of  integrator saturation  mentioned previously.  
Eighty-five percent of this light is emitted in a $20~\secs$ window centered on   the main fracture.
We next estimate the drop in elastic energy of the crystal during the main fracture.  
Neglecting 
the hole in the DCDC geometry, the elastic energy in the crystal of section $S$ (perpendicular to the force $F$) and of volume $V$ is $\left( \frac{F}{S}\right)^2 \times \frac{V}{2E} $, where $E$ is Young's modulus.
Applying a standard acoustic measurement of longitudinal and shear  waves~\cite{krautkramer_ultrasonic_1990} to BGO, we find $E= 139 \pm 4~\GPa$, somewhat above the  range of published values 
(up to $106 \pm 10~\GPa$~\cite{lebeau_monocrystalline_1985} and references therein).
Assuming the loaded volume and surface vary little, the drop in force of $\approx 5~\Newt$ is equivalent to a drop in elastic energy of $\approx 30~\microJoules$. 
Since the surface created during the fracture is roughly $20 \times 3 = 60~\mmsq$, this amounts to a surface creation energy of the order of $1~\Joules/\msq$, consistent with the scales typically 
found for crystals (e.g. $0.28~\Joules/\msq$ for~\baft, $1.2~\Joules/\msq$ for~\mgo~\cite{gilman_direct_1960}).
Furthermore, the conversion efficiency of elastic to luminous energy is therefore at least $6.5~\GeV/30~\microJoules \approx 3 \times 10^{-5}$.
A fuller understanding of the energy budget requires knowledge of the energy transformed into phonons, or heat.  The standard room-temperature technique to study brittle fracture only involves measurement of acoustic phonons using acoustic emission.  We are not aware of any work quantifying the fraction of energy going into acoustic phonons, or calibrating the energy scale of acoustic phonons.  The harder-to-implement technique of cryogenic calorimetry would allow measurement of all the phonons and would therefore provide more insight into the energy budget~\cite{astrom_brittle_2007}.

We have evidenced correlated mechanical-stress-induced emission of photons 
and of phonons 
in several common inorganic scintillators used for particle detection (\bgo, \zwo\ and \cdwo), at room temperature and in a normal atmosphere. 
At least for BGO, both emissions share similar distributions of waiting times over several decades.  Also for BGO, the two energy distributions are similar, and the wavelength of fractoluminescence is the same as that of scintillation.
In a mechanically-stressed scintillation-only detector, the power-law distribution of mechanoluminescence energies means that spurious low-energy events 
may mimick dark matter ones.  
It seems reasonable to assume our results apply to other tungstates, and it would be interesting to study the light-phonon correlations of individual fracture events in CRESST~II-like conditions of vacuum and low-temperature (other background-based explanations of the intriguing CRESST~II events have been proposed~\cite{kuzniak_surface_2012}).
In addition,
the re-absorption of products of fracto-emission
could  affect other types of solid-state particle detectors (e.g. ionization or ionization-phonon ones~\cite{knoll_radiation_2000,schnee_introduction_2011}).
From the standpoint of fracture physics, in contrast to previous studies 
by acoustic emission or by spectrally-resolved luminescence (e.g.~\cite{pallares_fractoluminescence_2012}), our use of the light channel  offers a precisely calibrated energy scale.  This allows us to quantify the amount of elastic energy converted into light, and is a step towards a better fundamental understanding of  fracture mechanisms.

\section{Acknowledgements}
This work has been funded in Canada by NSERC (Grant SAPIN 386432), CFI-LOF and ORF-SIF (project 24536), and by the France-Canada Research Fund (Project "Listening to Scintillating Fractures").
Prof.~C.~Dujardin of Lyon, France, kindly provided access to his equipment to measure X-ray emission spectra.
We thank M.~Chapellier for stimulating discussion,  and A.~B.~McDonald for comments on our manuscript.

\end{document}